\documentclass[twocolumn]{aastex631}
\usepackage{amsmath}
\usepackage{graphicx}
\usepackage{rotating}
\usepackage{txfonts}

\def\aap{{Astronomy and Astrophys.}}

\def\apj{{\it Astrophys. J.\ }}
\def\apjs{{\apj\ \it Suppl.\ }}		
\def\apjl{{\apj\ \it Lett.\ }}

\begin{document}
	
\nolinenumbers

\title{Flat Spectra of Energetic Particles in Interplanetary Shock Precursors }

\author{Mikhail Malkov}

\affiliation{Department of Astronomy and Astrophysics, University of California, San Diego, La Jolla, CA 92093, USA}

\author{Joe Giacalone}

\affiliation{Lunar \& Planetary Laboratory, University of Arizona, Tucson, AZ
85721, USA}

\author{Fan Guo}

\affiliation{Los Alamos National Laboratory, Los Alamos, NM 87545, USA}
\begin{abstract}
The observed energy spectra of accelerated particles at interplanetary
shocks often do not match the diffusive shock acceleration (DSA) theory
predictions. In some cases, the particle flux forms a plateau over
a wide range of energies, extending upstream of the shock for up to
seven flux's e-folds before submerging into the background spectrum.
Remarkably, at and behind the shock that we have studied in detail, the
flux falls off in energy as $\epsilon^{-1}$, consistent with the
DSA prediction for a strong shock. The upstream plateau suggests a
different particle transport mechanism than those traditionally employed
in DSA models. We show that a standard (linear) DSA
solution based on a widely accepted diffusive particle transport with
an underlying resonant wave-particle interaction is inconsistent with
the plateau in the particle flux. To resolve this contradiction, we
modify the DSA theory in two ways. First, we include a dependence
of the particle diffusivity $\kappa$ on the particle flux $F$ (nonlinear
particle transport). Second, we invoke short-scale magnetic perturbations
that are self-consistently generated by, but not resonant with, accelerated
particles. They lead to the particle diffusivity increasing with the
particle energy as $\propto\epsilon^{3/2}$ that simultaneously decreases
with the particle flux as $1/F$. The combination of these two trends
results in the flat spectrum upstream. 
\end{abstract}

\section{Introduction}

Diffusive shock acceleration (DSA) \citep{krym77,axf78,Bell78,BlandOst78}
is arguably the most universal and robust mechanism whereby particles
can be accelerated to high energies in shocks across the universe.
Its physical and intuitive grounds are comprehensible. The particle
momentum spectrum behind the shock comes from a ``back-of-the-envelope''
calculation. It is a power-law $\propto p^{-q}$, with an index $q=3r/\left(r-1\right)$
that, to the first approximation, depends only on the shock compression
$r$.

The simplicity of the DSA is, however, deceptive. After almost half
a century of research, it is still challenging to calculate the rate
at which it operates, and the maximum energy particles can gain in
realistic shock environments. Its index $q$ often disagrees with
observations even if the shock is known to be strong, and the index
$q$ should be equal to four. When the spectrum is harder than predicted,
the difference is usually explained by a nonlinear shock modification
due to the growing pressure of accelerated particles (see \cite{MDru01}
for a review). When softer, the disagreement can, e.g., be attributed
to the non-stationarity and curvature of the shock, strong short-scale
magnetic perturbations generated by accelerated particles, and propagation of particle scattering waves relative to the plasma flow \citep{Kennel1986,Bell_2019MNRAS,Hanusch_2019,MalkovAharonian2019,Diesing2021}.
These arguments are primarily applied to the supernova remnant shocks,
in which measurements of energetic particles are limited in accuracy
and indirect by nature, thus obscuring the cause of the spectral softening.
 The particle spectra are inferred from the emission of accelerated
electrons and, maybe, from the $\gamma$- radiation generated by accelerated
protons interacting with adjacent molecular clouds, if present. If
so, telling the radiatively more efficient leptons from overwhelmingly
more abundant hadrons is still challenging.

This paper considers an even more puzzling DSA disagreement with the
observed spectra. Notably at some, but not all, interplanetary shocks\emph{
}observed \emph{in situ, }e.g.,\emph{ }\citep{Lario_2018,Lario_2022,Perri2023},
the particle flux \emph{flattens} \emph{upstream,} whereas the downstream
part still agrees with the DSA. Since the disagreement is partial,
it helps identify the DSA elements responsible. In addition, it might
shed light on how the DSA is sped up by waves excited by the accelerated
particles themselves \citep{Bell78}. Recall that within the DSA,
particles gain energy when they cross and recross the shock by scattering
off magnetic perturbations of whatever origin. Therefore, strong self-generated
waves ``bootstrap'' the particle acceleration.

Central to the bootstrap is a simultaneous growth of wave amplitudes
and their lengths during the acceleration. Scattering is most efficient
when the wave-particle resonance condition, $kr_{\text{g}}\left(p\right)\sim1$,
is maintained throughout the wave and particle spectra up to at least
the maximum particle momentum, $p_{\text{max}}$. Here, $r_{\text{g}}$
is the particle gyroradius, $r_{g}=cp/eB_{0}$, and $k$ is the wave
number of resonant Alfven waves. However, the fastest-growing waves
do not necessarily scatter particles most rapidly, which is required
for efficient acceleration.

Resonant waves typically saturate at a level not significantly higher
than $\delta B/B_{0}\sim1$ \citep{Voelk84}. Macroscopically-driven
nonresonant instabilities may continue to grow beyond this level.
Two types of them are often invoked in DSA treatments. One is the
current-driven, also called Bell's instability \citep{Bell04}. The
other one is an acoustic instability driven by the pressure gradient
of accelerated particles \citep{DruryFal86}. We will argue that a
\emph{nonresonant} wave-particle interaction may be critical to the
spectrum flattening observed ahead of interplanetary shocks. At the
same time, it does not affect the DSA-predicted spectral slope downstream,
as also observed.

We have organized this paper along its line of arguments as follows.
Because the flat upstream spectra are uncommon, in the next section
we specify particle transport regimes that may result in such spectra.
Next, we review the turbulence spectra that may be consistent with
such regimes. Each of these two steps will be made assuming two different
types of wave-particle interaction: resonant and nonresonant. The
above-described analyses will allow us to eliminate some common combinations
of turbulence spectra, particle transport regimes, and wave-particle
interaction types (resonant vs nonresonant) inconsistent with the
flat spectra. In Sec.\ref{subsec:Turbulence} we show that the wave-particle
interaction is likely nonresonant, the particle transport must be
nonlinear (flux dependent), and the wave spectrum is likely to be
of the Iroshnikov-Krainchnan type. In Sec.\ref{sec:Particle-Transport-At-Shock}
we discuss a nonlinear transport regime that likely leads to the flat
spectra. Sec.\ref{sec:Acceleration} deals with the acceleration model
based on this transport regime. In Sec.\ref{sec:Fitting-the-Spectra},
we fit the model predictions to the data. The paper concludes with
a summary and discussion of its results.

\section{Observational Hints \label{sec:Observational-Hints}}

In this section, we constrain the particle transport and its underlying
MHD turbulence required for the spectral flattening observed at the
interplanetary shock of May 2005, e.g., \citep{Lario_2018,Lario_2022,Perri2023}.
We selected this data set because the flat particle flux upstream,
inconsistent with the DSA, coexists with its DSA-compliant downstream
counterpart. This example offers insights into conditions under which
the "standard" (linear diffusion) DSA model
fails. It helps eliminate turbulence
and transport combinations inconsistent with the data. We will show
that if the particle diffusivity does not depend on their intensity,
it is difficult to explain the flat particle flux. We also rule out
several types of turbulence spectra. These analyses help us zero in
on a unique combination of turbulence and particle transport regimes
leading to the flat particle flux upstream of the shock.

\subsection{Constraining Particle Diffusion Coefficient\label{subsec:Diffusivity}}

\begin{figure*}
	\centering
\includegraphics[scale=0.3]{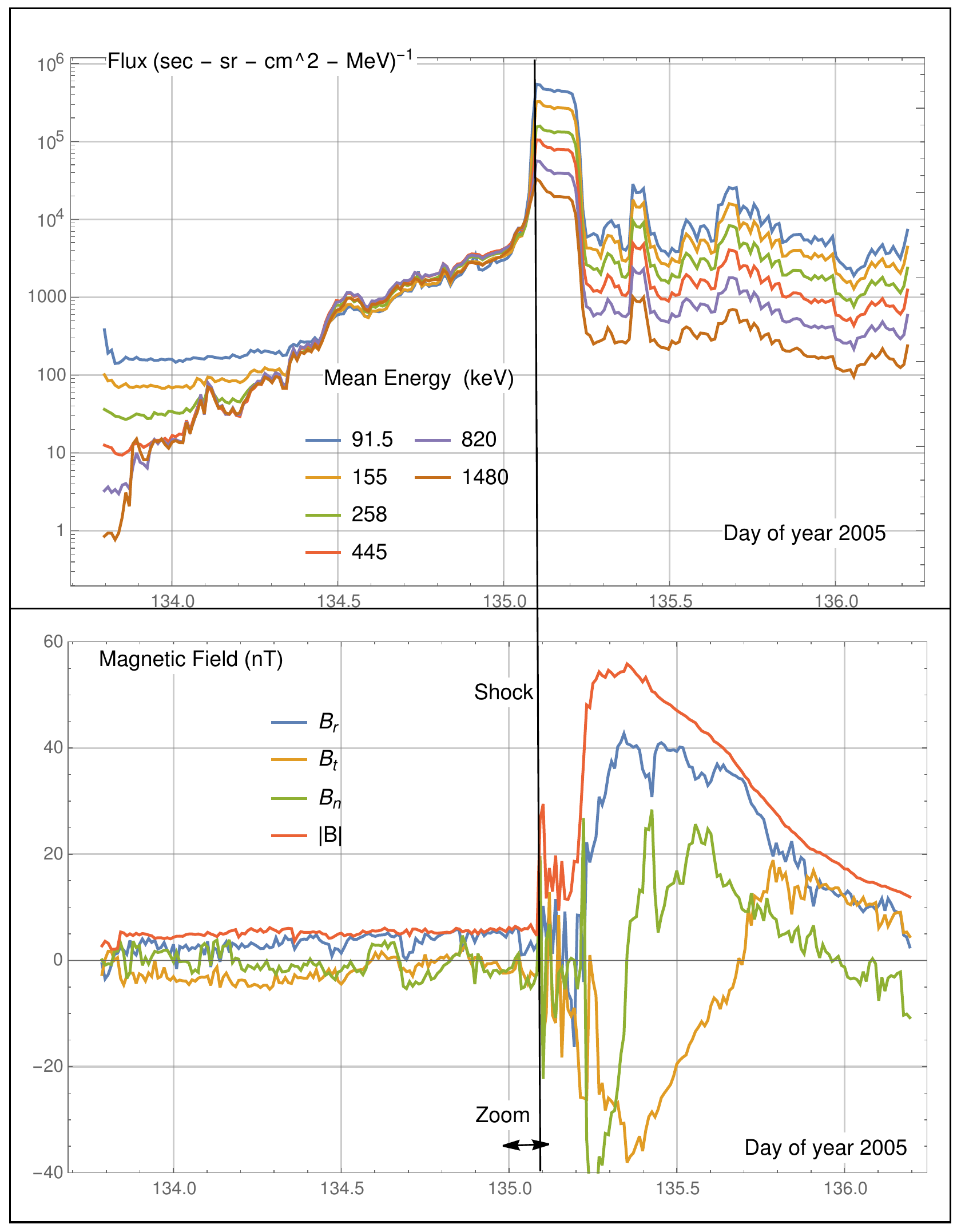}\includegraphics[scale=0.3]{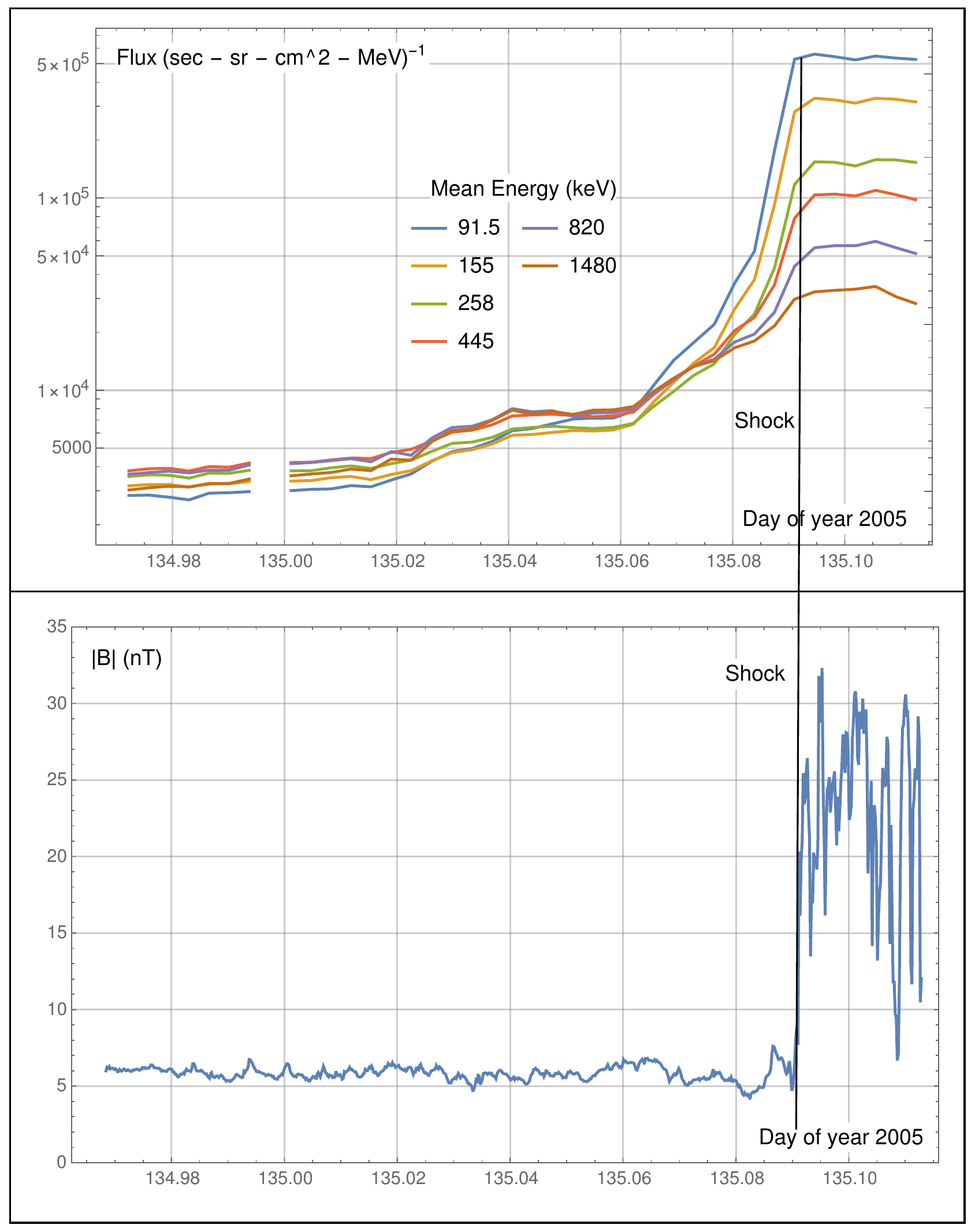}
\caption{Top panels: EPAM 5-minute averaged solar particle fluxes. Bottom panels:
MAG 16-second averaged interplanetary magnetic field data. The right
two panels zoom into a near shock region, indicated in the bottom
left panel. The data downloaded from the ACE Science Center \protect\protect\url{http://www.srl.caltech.edu/ACE/ASC/level2}
\label{fig:ObservSpectra}}
\end{figure*}

Figure \ref{fig:ObservSpectra} demonstrates disagreements with the
"standard" DSA model. However, we start
with what is agreeable. On the downstream side, the particle flux
decreases approximately as $\epsilon^{-1}$ with energy, consistent
with the DSA prediction for a shock with compression ratio close to
four. Immediately on the upstream side, the low-energy part of the
spectrum initially decays more steeply with distance from the shock.
This behavior is also \emph{qualitatively} consistent with the DSA
if the particle diffusivity grows with energy.

Further upstream, the disagreements with the DSA become obvious. Let
us consider a general steady-state DSA solution in a fixed scattering
environment. As we mentioned earlier, the particle scattering can
be enhanced by unstable magnetic perturbations. However, in the most
DSA schemes, this enhancement does not change the slope of the spectrum.
It decreases the acceleration time. The scattering perturbations may
saturate at a fixed $\delta B\sim B_{0}$ level if the particle intensity
is sufficient to drive them to this level. Assuming also that the
scattering supports particle diffusion with a particle diffusivity
$\kappa\left(\epsilon,z\right)$, we first examine which $\kappa$
might make the spectrum flat at some distance upstream, provided that
the spectrum is $\epsilon^{-s}$ at and behind the shock, as observed.

Upstream particles conserve their energy as long as the second-order
Fermi acceleration and the shock modification by accelerated particles
are negligible. The basic DSA solution arises from a balance between
convective and diffusive particle fluxes. No particle losses are assumed;
particles do not escape from those parts of the shock precursor where
the balance is maintained. In the shock reference frame, moving at
a speed $u,$ on its upstream ($z<0$) side, we thus can write:

\begin{equation}
\kappa\left(\epsilon,z\right)\frac{\partial F}{\partial z}-uF=0.\label{eq:CDbalance}
\end{equation}
To compare the DSA solution with the data, we use here the particle
flux, $F\left(\epsilon\right)d\epsilon=f\left(p\right)vp^{2}dp$,
with energy normalization, instead of the particle distribution, $f$,
normalized to $4\pi p^{2}dp$. Here $v=p/m$ is the particle velocity. 
According to the DSA, particles injected
at $\epsilon\sim mu^{2}$ at the shock front develop the spectrum
$F=Q_{0}\epsilon^{-s}$ for $\epsilon\gg mu^{2}$, where $Q_{0}$
is the intensity of seed particles extracted from the thermal pool
at the shock interface. Here the index $s=\left(r+2\right)/2\left(r-1\right)=q/2-1$,
and $r$ is the shock compression. The transition of $F\left(\epsilon\right)$
between the thermal downstream core and the power-law part of the
spectrum, $\propto\epsilon^{-s}$, can also be calculated analytically,
given the seed extraction mechanism at $\epsilon\sim mu^{2}$ (see
\cite{MDru01} and references therein). In what follows, we focus
on sufficiently high energies where the power-law dependence is established.

The data shown in Fig.\ref{fig:ObservSpectra} indicate that $s\approx1$,
as predicted by the DSA for strong shocks. If $s$ deviates from the
value shown above, the balance in Eq.(\ref{eq:CDbalance}) is likely
to be violated (see, e.g., Sec.\ref{sec:Acceleration} below). This
equation also implies that the upstream plasma, inflowing into the
shock at the speed $u,$ carries no pre-accelerated particles from
$z=-\infty$ ($F\to0$), and eq.(\ref{eq:CDbalance}) remains valid
up to $z=-\infty$. Not being fully consistent with the data, this
assumption is not essential as long as $\kappa$ does not depend on
$F$, while both $F$ and $\partial F/\partial z$ are negligible
far upstream.

Far upstream, the flux in Fig.\ref{fig:ObservSpectra} flattens abruptly
in $z$ in each energy channel, proceeding from lower to higher energy.
It almost certainly manifests a preexisting background spectrum. Since
it is much lower than the flux in the area of the disagreement with
the DSA, we have neglected it in eq.(\ref{eq:CDbalance}), but will
use it as a boundary condition when fitting the data in Sec.\ref{sec:Fitting-the-Spectra}.
Assuming here that all the accelerated particles are initially injected
at the shock front from the plasma thermal core, a complete solution
at $\epsilon\gg mu^{2}$, upstream and downstream, can be written
as:

\begin{equation}
F=\begin{cases}
Q_{0}\epsilon^{-s}\exp\left[-u\int_{z}^{0}\frac{dz}{\kappa\left(\epsilon,z\right)}\right], & z<0\\
Q_{0}\epsilon^{-s}, & z\ge0
\end{cases}\label{eq:fSimpleSol}
\end{equation}

Immediately behind the shock, the observed spectrum remains nearly
constant (see zoom in Fig.\ref{fig:ObservSpectra}), which justifies
the above solution at $z\ge0$. However, the spectrum gradually decays
at larger $z$. This decay contrasts evolving curved shocks with stationary
plane ones, which is seen from an analytically solvable case of the
DSA at expanding spherical shocks \citep{MalkovAharonian2019}. Further
downstream (at $t\approx135.2$), the flux drops sharply, presumably
because it crosses a magnetic ``piston''. It is associated with
a CME that likely drives the shock. We will not further discuss these
two downstream flux decreases as they are unlikely to affect the upstream
flattening.

Now, we test if the spectrum can be flat in an extended region upstream
for any realistic $\kappa$ in eq.(\ref{eq:fSimpleSol}). If $\kappa$
grows with the particle energy and is $z-$ independent, the energy
spectrum has an isolated maximum at some $\epsilon_{\text{max}}\left(z\right)$.
It can be obtained from the equation $\kappa^{2}\left(\epsilon\right)/\epsilon\kappa^{\prime}\left(\epsilon\right)=u\left|z\right|/s$.
The spectrum is, therefore, not flat in any extended area upstream.
Note that the particle flux decreases in the flat spectrum area by
up to three orders of magnitude (depending on the energy). Therefore,
we can safely reject the above scenario, in which $\kappa$ is $z$
-independent everywhere upstream.

Based on the upper right panel of Fig.\ref{fig:ObservSpectra}, the
spectrum flattens at $z<z_{0}$, where $z_{0}$ roughly corresponds
to $t=135.06$. If the solution in eq.(\ref{eq:fSimpleSol}) applies,
$F$ must be energy independent at $z=z_{0}$, but it still must decrease
as $\epsilon^{-s}$ at $z=0$. From eq.(\ref{eq:fSimpleSol}) we thus
have

\begin{equation}
u\int_{z_{0}}^{0}\frac{dz}{\kappa\left(\epsilon,z\right)}=s\ln\frac{\epsilon_{0}}{\epsilon},\label{eq:KappaLnEps}
\end{equation}
where $\epsilon_{0}>\epsilon$ relates to the normalization of $F$.
Beyond $z_{0}$ the spectrum is observed to be flat. We can infer
the transition point $z_{0}$ from the observations, which is about
the same for all energy channels. The above relation means that the
quantity $\overline{\kappa^{-1}}$ ($z$-averaged over $z_{0}<z<0$)
must decrease with energy as $\ln\left(\epsilon_{0}/\epsilon\right)$.
This relation does not, however, fully constrain the turbulence spectrum
in the area $z_{0}<z<0$. We will return to that constraint later.

The spectral flatness at $z<z_{0}$, requires $\kappa$ to be $\epsilon$-independent
in this area. To examine whether physically reasonable wave spectra
can meet this requirement and eq.(\ref{eq:KappaLnEps}), let us start
with the most straightforward possibility, assuming that the particle
diffusion is field-aligned. It is a good approximation if the wave
amplitudes are moderate and the angle between the shock normal and
magnetic field, $\vartheta_{Bn}$, is not too close to $\pi/2$: $\delta B^{2}/B_{0}^{2}<\cot\vartheta_{Bn}$
(Sec.\ref{sec:Particle-Transport-At-Shock}). A common DSA assumption
is that the particle cyclotron resonance with waves supports their
diffusion, so $k_{\parallel}\approx\omega_{c}/v_{\parallel}$, where
$v_{\parallel}$ is the particle velocity along the field. It is assumed
that the wave frequency $\omega\ll\omega_{c}$. Hence, for the diffusion
coefficient, we have, $\kappa\sim\kappa_{\parallel}\propto v^{3}/\delta B_{k}^{2}$,
e.g., \citep{Lee82}. This relation links $z$ with $\epsilon$ by
virtue of eq.(\ref{eq:KappaLnEps}) because $\delta B_{k}$ depends
on $z$. Namely, it requires $z$- averaged magnetic fluctuation spectrum
to be $\overline{\delta B_{k}^{2}}\propto k^{-3}\ln\left(k/k_{0}\right)$
within $z_{0}<z<0$, where $k_{0}=\omega_{c}/\sqrt{2\epsilon_{0}/m}$.
Since $\kappa\propto v^{3}/\delta B_{k}^{2}$, it becomes energy independent
if we neglect the logarithmic dependence on $k$ of $\overline{\delta B_{k}^{2}}$,
or assume that the $z$- averaging somehow compensates for this insignificant
factor, compared to $k^{-3}$, in the spectral density. These are
relatively mild changes, though. For $z<z_{0}$, one thus obtains:
$\delta B_{k}^{2}\propto k^{-3}$, which makes $\kappa$ roughly energy
independent, as required for the flat spectrum. Let us now consider
the possibility of generating the $k^{-3}$ spectrum.

\subsection{Constraining the Underlying Wave Turbulence\label{subsec:Turbulence} }

The turbulence spectrum often discussed as being resonantly generated
by accelerated particles is the $k^{-2}$ spectrum, associated with
nonrelativistic particles \citep{Lee1983,Forman1985} (and $k^{-1}$
with the relativistic ones \citep{Bell78}). Both cases are valid
for $r=4$ shock compression ratio, thus corresponding to $p^{-4}$
spectrum downstream (normalized to $p^{2}dp$). Resonant wave-particle
interactions without wave cascading are employed in obtaining these
results. The steep $k^{-3}$ wave spectrum, inferred in the preceding
subsection, would correspond to a very flat, $p^{-3}$ particle spectrum
at and behind the shock. The latter is hardly possible as even in
the limit of much more powerful shocks, strongly modified by particles
accelerated to ultra-relativistic energies, the flattest asymptotic
downstream spectrum is $p^{-3.5}$ \citep{MDru01}. However, the spectrum
in question, corresponding to $F\left(\epsilon\right)=const$, is
observed farther upstream. As we mentioned, Lee's theory relates the
wave spectrum upstream to the particle spectrum downstream, which
is an oversimplification in describing the flat upstream spectra.

According to Fig.\ref{fig:ObservSpectra}, the spectrum remains flat
despite decaying by 2-3 orders of magnitude upstream. This behavior
is counterintuitive and defies the standard DSA principles. Indeed,
particles observed far upstream have likely spent most of their last
acceleration cycle (crossing and recrossing the shock gaining $\sim U_{\text{sh}}/v$
of their current energy) while balancing between the convection with
the flow toward the shock and diffusing against it. Diffusion generally
intensifies with the particle energy, which should lead higher energy
particles to diffuse farther upstream. It seems, they should break
the spectrum flatness, whatever mechanism maintains it up to a certain
distance. On the contrary, the spectrum remains flat in a wide area
upstream.

Unstable waves, initiated far upstream, where the particle flux is
low, are convected with the flow to the shock, and their growthrate
must increase along with the particle flux. The turbulence intensifies
if no significant instability suppression (quasi-linear or nonlinear)
occurs. Eventually, it may saturate and change its spectral shape
before a fluid element crosses the shock. The question is why and
how the particle spectrum remains flat, whereas the transport driving
turbulence evolves. To answer this question, we consider the following
scenario.

When the wave and particle flux grow toward the shock, the particle
transport goes nonlinear, thus making their diffusivity, $\kappa$,
flux-dependent \citep{Bell78,Lee82}. Meanwhile, the resonant wave-particle
interaction, that is behind the quasi-linear calculations of particle
transport in the above references and Sec.\ref{subsec:Diffusivity},
ceases to apply if the wave amplitudes grow beyond the level of a
resonant, quasi-linear regime of the wave-particle interaction. Assuming
further that the waves are nonresonantly driven by a pressure gradient
of energetic particles \citep{DruryFal86}, we may use some results
of \citep{MalkMosk_2021}. According to them, an ensemble of weak
shocks (shocklets) generated by the pressure gradient evolves toward
longer scales by shock mergers (inverse turbulence cascade). As their
strength increases, the spectrum develops a magnetic turbulence component
transferring to shorter scales (forward cascade). The magnetic part
of the turbulence supports particle scattering, thus controlling their
spatial transport. The forward cascade flattens the spectrum to a
$k^{-3/2}$ Iroshnikov-Kraichnan (IK) spectrum \citep{Iroshnikov1964,Kraichnan1965}.

The IK spectrum is not a prerequisite for flat particle flux, as shown
in the sequel. Nonetheless, the authors of \citep{Perri2023} have
identified the wave spectrum $k^{-1.51}$ precisely in the area of
the flat particle spectrum (see Fig.4 in their paper). Compared to
the canonical DSA spectrum ($k^{-2}$ in a nonrelativistic regime),
the IK spectrum contains more energy in shorter scales. Given their
significant amplitudes, a particle sufferes many scattering events
while interacting with a broad range of waves, still much shorter
than its Larmor radius. In effect, it accumulates a significant deflection
before completing its Larmor rotation.

By contrast, most of the DSA studies rely on resonant wave-particle
interactions, assuming a low-amplitude, random-phase wave spectrum
with $\delta B_{k}\ll B_{0}$. In some cases, though not frequently,
an opposite approximation of a single dominant wave with $\delta B_{1}\gtrsim B_{0},$
in a broad quasi-linear $\delta B_{k}\ll B_{0}$ continuum compares
positively with simulations and even observations (see, e.g., \cite{Hanusch2019ApJ}
and references therein). These approaches do not work for the case
at hand, both\emph{ de facto } and because the required conditions
do not hold. As seen from Fig.\ref{fig:ObservSpectra}, separate field
components reach the level $\delta B\gtrsim B_{0}$ and often form
coherent short-scale structures. If the spectrum is sufficiently flat
and intense, particles with a large Larmor radius should be deflected
by such perturbations multiple times during each rotation. For sufficiently
large amplitudes, an approximation based on a sequence of uncorrelated
particle deflections by the magnetic perturbations with the scales
$l\ll r_{g}$ but large amplitude $\delta B_{l}\gtrsim B_{0}$ has
proved more accurate.

As the IK spectrum is significantly flatter than Lee's resonant $k^{-2}$
spectrum, we may assume that the wave-particle interaction with high-energy
particles is dominated by a $kr_{g}\gg1$ condition, that is essentially
nonresonant. At the same time, the phases of the short-scale magnetic
perturbations are randomized as the waves cascade to shorter scales.
The nonresonant transport is then easily calculated in the above-described,
well-known fashion. One starts with an angular diffusion in momentum
space, assuming that by crossing the turbulence correlation length,
$l$, particles deflect only by a small angle $\delta\vartheta$.
For the angular diffusion rate we find $\nu=\Delta\vartheta^{2}/2t\sim\left(lv/r_{g}^{2}\right)\left(\delta B_{l}/B_{0}\right)^{2}$.
Here, we assumed that the angle $\Delta\vartheta$ is accumulated
from $vt/l\gg1$ uncorrelated deflections, each of which at an angle
$\delta\vartheta\sim\left(l/r_{g}\right)\left(\delta B_{l}/B_{0}\right)$.
Note that unlike the standard quasi-linear derivation that we applied
to the resonant wave-particle interaction leading to the parallel
diffusion $\kappa_{\parallel}\propto v^{3}/\delta B_{k}^{2}$, the
length scale $l$ in the amplitude $\delta B_{l}$ is not related
to the particle velocity, $v$. Thus, we arrive at the following expression
for $\kappa_{\parallel}$:

\begin{equation}
\kappa_{\parallel}=\frac{v^{2}}{3\nu}\text{\ensuremath{\sim\frac{v^{3}}{l\omega_{c}^{2}}\left(\frac{B_{0}}{\delta B_{l}}\right)^{2}}}\label{eq:kappa-par}
\end{equation}
Since $l$ is a fixed turbulence correlation length, it is not associated
with the resonant wave number $k=r_{g}^{-1}\propto1/\sqrt{\epsilon}$;
the parallel diffusivity scales with energy as $\kappa_{\parallel}\propto\epsilon{}^{3/2}$,
instead of $\kappa_{\parallel}\propto\epsilon^{3/4}$, which one would
obtain in the case of a resonant particle diffusion, specifically
for the IK spectrum.

\section{Possible Regimes of Particle Self-Confinement\label{sec:Particle-Transport-At-Shock}}

To fit the data shown in Fig.\ref{fig:ObservSpectra}, in this section,
we introduce some further modifications to the ``standard'' DSA
mechanism. We also justify simplifications, such as a planar and stationary
shock assumption that we will use later. Although the diffusion-convection
balance essentially controls particle transport upstream, similarly
to eq.(\ref{eq:CDbalance}), we now also allow for an injection of
thermal particles at the shock discontinuity, escape of accelerated
and influx of preexisting energetic particles from the far upstream
space. Therefore, the total particle flux, diffusive plus convective,
across the shock precursor is not exactly zero, although it is much
smaller than its primary components.

Freshly injected particles are vital for spectral flatness. Indeed,
some four orders of magnitude enhancement of the particle flux at
the shock, compared to the background ($z=-\infty$), would be impossible
by a mere reacceleration of the background population, preexisting
far upstream. The reaccelerated flux enhancement would hardly exceed
a factor of a few, depending on the shock compression and the background
upstream spectrum. This aspect of the DSA was discussed in detail
by \citep{MalkMosk_2021} in conjunction with a recently discovered
fine structure in a galactic cosmic ray spectrum, which also deems
incompatible with the ``standard'' DSA.

On the far upstream end of the flat spectrum region, the simple diffusion-convection
balance in eq.(\ref{eq:CDbalance}) needs to be supplemented with
the background CRs convected into the shock precursor and shock-accelerated
CRs diffusively leaked from it. Inside the flat spectrum region and
in the shock vicinity, the particle flux is greatly enhanced over
the background level, which couples it with the wave intensity. By
contrast, a simple relation between the particle flux and wave intensity
is not accurate in the far upstream region, where both quantities
approach their background levels and decouple. Their intensities are
generally unrelated to each other. Therefore, a linear relation between
the particle flux and wave energy density introduced below by eq.(\ref{eq:Ew-of-P})
contains a model parameter, $\psi$. It is, however, essential only
in the transition region to the background energetic particles and
does not significantly affect most of the shock precursor where the
flat particle flux is observed.

The spatial downstream distribution of shock-accelerated particles
is qualitatively sensitive to even a gradual time dependence of the
shock speed and its front curvature. In a steady planar shock, the
downstream distribution is homogeneous. If the shock slows down and
expands in a self-similar way, e.g., after a point explosion, the
distribution of accelerated particles is more complicated but can
still be obtained from a self-similar solution \citep{MalkovAharonian2019}.
The solution shows that the particle distribution decays behind the
shock as well. This self-similar solution is relevant to the downstream
particle distribution shown in Fig.\ref{fig:ObservSpectra} because
CME shocks in the heliosphere are also curved and decelerating. However,
the gradual flux decrease shown in Fig.\ref{fig:ObservSpectra} and
interpreted in the above sense is not critical for the upstream flattening.
Hence, the stationary and planar shock approximation with a constant
flux downstream in studying the upstream flattening appears justified.

The shock under consideration has been found to be oblique with $\vartheta_{Bn}\approx62^{\circ}$
\citep{Perri2023}. So, we also need to estimate the effect of transport
anisotropy. It is characterized by two components of the diffusion
tensor, $\kappa_{\parallel}$ and $\kappa_{\perp}$, along and across
the magnetic field, respectively. Their combination governs the diffusion
along the shock normal, $\kappa=\kappa_{\parallel}\cos^{2}\vartheta_{Bn}+\kappa_{\perp}\sin^{2}\vartheta_{Bn}$.
Since the value $\tan^{2}\left(\vartheta_{Bn}\right)\approx3.5$ is
relatively large, $\kappa_{\perp}$ may significantly contribute to
the particle diffusion immediately ahead of the shock, where it is
enhanced and $\kappa_{\parallel}$ is suppressed by a self-driven
turbulence. However, $\kappa_{\perp}\ll\kappa_{\parallel}$ where
the magnetic field fluctuations remain limited $\delta B\lesssim B_{0}$,
as we discussed at the end of Sec.\ref{subsec:Diffusivity}. Therefore,
$\kappa_{\perp}\sim\kappa_{\parallel}$ only in a narrow region ahead
of the shock. Farther upstream, where the spectrum flattens, the cross-field
diffusion decreases, since the wave energy $E_{\text{w}}$ does, while
$\kappa_{\perp}\sim\kappa_{\parallel}E_{\text{w}}^{2}$ (the dimensionless
wave energy $E_{\text{w}}$, normalized to the baground magnetic field
energy, is introduced below). A rough estimate of $E_{\text{w}}$
in the upstream area adjacent to the shock can be made by assuming
that the wave pondermotive pressure exerted on particles is equilibrated
by their partial pressure (see eq.(\ref{eq:Ew-of-P}) below). It shows
that $E_{\text{w}}\sim$$\sqrt{\epsilon/m}\epsilon F\left(\epsilon\right)/n_{0}V_{A}^{2}$.
Here $n_{0}=\rho/m$ is the plasma density. By scaling $E_{\text{w}}$
down from its maximum value at $\sqrt{\epsilon/m}\sim10^{9}$cm/s,
assuming $n_{0}\sim1$ cm$^{-3}$, $V_{A}\sim10^{7}$ cm/s, we can
place the following upper limit $E_{\text{w}}\lesssim10^{-5}\epsilon F\left(\epsilon,z\right)$.
By noting that $\epsilon F\left(\epsilon,z=0\right)\approx const\approx5\times10^{5}$
(Fig.\ref{fig:ObservSpectra}), while sharply decreasing at $z<0$,
we conclude that the quasilinear treatment of the wave interactions
with plasma we invoke below in the flat region, is tenable.

In the flat spectrum region, the $\vartheta_{Bn}$ appears to decrease
compared to the near upstream, even though with considerable variations.
These trends can be gleaned from Fig.1 in \citep{Perri2023}. Therefore,
$\kappa_{\perp}$ might significantly contribute to the particle diffusion
upstream only in a relatively narrow transition zone between the shock
and the flat spectrum area, where $E_{\text{w}}$ attains its maximum
at $E_{\text{w}}\sim5$. In this transition zone, low-energy particles
exhibit a steeper flux decay, which is necessary to converge to the
flat particle flux farther upstream. In understanding the nature of
this spatial decay, one needs to consider the angular particle dynamics
in momentum space underlying the diffusive approximation. As we see
from Fig.\ref{fig:ObservSpectra} (bottom right), the $\left|B\right|$
-profile makes traps and magnetic barriers for the particles immediately
ahead of the shock. It is reasonable to assume that the trap adjacent
to the shock is filled with the downstream $\epsilon^{-1}$ particles,
leaking in the upstream direction and forming the flat spectrum. We
can conjecture that these magnetic structures regulate the particle
leakage, making their flux almost energy-independent, at least in
this particular case of the May 2005 shock.

Consider such a magnetic trap near the shock front. Its loss-cone
angle is defined by the trap's mirror ratio, $B_{\text{min}}/B_{\text{max}}$
so that only particles that have pitch angles at the bottom of the
trap $\sin^{2}$$\alpha<B_{\text{min}}/B_{\text{max}}$ may leak through
the trap's barrier where $B$ approaches $B_{\text{max}}$. Adiabaticity
of particle motion is assumed but increasingly violated for higher
energies, so these particles are more likely to leak, thus facilitating
the formation of energy-independent flux upstream. The energy-biased
leakage mechanism warrants a separate study, which is beyond the scope
of this paper. An observationally-learned fact is that the leaking
particle flux becomes and remains flat farther upstream of the trap.
The flat flux sustainability is the focal point of this study.

On a practical note, neglecting $\kappa_{\perp}$ allows us to solve
eq.(\ref{eq:FiOfP1}) below in explicit form. With $\kappa_{\perp}$
included, it can be solved implicitly via an inverse function $z\left(F,p\right)$,
which would only obscure the interpretation of the result but give
us no further insight into the physics of the spectral flattening.
As we will see, the parallel diffusion alone supports a rapid transition
of the power-law spectrum at the shock, which is close to $\epsilon^{-1}$,
into the flat spectrum upstream. Therefore, neglecting the cross-field
diffusion in a narrow layer adjacent to the shock, where $E_{\text{w}}$
approaches unity and possibly even exceeds it, appears commensurate
with this level of consideration.

For the field-aligned diffusion, we follow Bell's approach to its
quasi-linear suppression by Alfven waves generated by accelerated
particles upstream \citep{Bell78}, except the wave-particle interaction
is nonresonant in our case. We define the dimensionless wave spectral
density, $E_{\text{w}}$, already used above, by relating it to the
RMS magnetic field fluctuations, $\left\langle \delta B^{2}\right\rangle $,
and the background field, $B_{0}$, as follows:

\[
\frac{\left\langle \delta B^{2}\right\rangle }{B_{0}^{2}}=\int E_{\text{w}}\left(k\right)d\ln k=\int E_{\text{w}}\left(p\right)d\ln p.
\]
The last relation conveniently implies an inverse dependence $p\propto k^{-1}$
\citep{Skill75a}, although we do not impose the resonance relation
$\rho_{g}\left(p\right)k\sim1$. Note that we use the same notation
$E_{\text{w}}$ for the $p$- and $k$- dependent spectral densities,
which should not lead to confusion. We also use physical arguments
of \citep{BlandPerturbCRShock80,Drury83} about the relation between
the wave generation by the work done by the pressure of accelerated
particles on the fluid. For that purpose, we introduce a dimensionless
partial pressure of energetic particles and normalize it, as $E_{\text{w}}$,
to $d\ln p$:

\begin{equation}
P\left(p\right)=\frac{8\pi}{3\rho V_{A}^{2}}vp^{4}f,\label{eq:PartPressDef}
\end{equation}
Here $f$ is the ordinary particle distribution function normalized
to $4\pi p^{2}dp$. The wave generation rate upstream of the shock
can be obtained from the rate at which the pressure of energetic particles
does work on the waves. The waves are propagating oppositely to the
inflowing plasma, so their speed in the shock frame is $u-V_{\text{w}}$,
with $V_{\text{w}}\approx V_{A}$. Balancing the wave pondermotive
pressure with that of the energetic particles, e.g., \citep{Drury83},
we have

\begin{equation}
\left(M_{A}-1\right)\frac{\partial E_{\text{w}}}{\partial z}=\frac{\partial P}{\partial z},\label{eq:WaveGen}
\end{equation}
where $M_{A}=u/V_{A}$ is the shock Alfven Mach number. This equation extends the pressure balance principle to a detailed balance that, generally speaking, requires the wave-particle resonance that we do not assume. However, the balance is crucial for the coordinate dependence of $P$ and $E_{\text{w}}$, not on momentum and wave number. 
For a successful
fit of the particle spectrum in the entire upstream region, it is also
essential that $E_{\text{w}}$ and $P$ decouple far from the shock
front, where they approach the respective background levels. Thus,
after integration, we ought to write:

\begin{equation}
E_{w}=\frac{16\pi\sqrt{2m}\epsilon^{3/2}}{3\left(M_{A}-1\right)\rho V_{A}^{2}}\left(F+\psi\right).\label{eq:Ew-of-P}
\end{equation}
Here $\psi\left(p\right)$ is an arbitrary function of particle momentum,
mentioned earlier, and $F\left(\epsilon,z\right)d\epsilon=vp^{2}f\left(p\right)dp$
is the particle flux normalized to $Fd\epsilon$, as introduced in
eq.(\ref{eq:CDbalance}). We will use this quantity instead of $f$
and $P$ for fitting the solution to the data. It might appear plausible
to obtain $\psi$ from the far upstream values of $E_{\text{w}}$
and $F$, i.e., at $z=-\infty$. However, the underlying eq.(\ref{eq:WaveGen})
is valid in a strongly nonlinear particle transport regime in which
both particle and wave intensities are high; other terms become essential
when the particle pressure gradient decreases to the background level.
Additional wave processes other than their convection and generation
by the particle pressure in eq.(\ref{eq:WaveGen}) need to be included,
such as linear damping, wave steepening, and nonlinear Landau damping.
Thus, $\psi\left(p\right)$ remains undetermined at this level of
the model. However, as we will see, $\psi$ does not affect the fit
in most of the shock precursor and is essential only at the transition
to the spatially independent background particle flux far upstream.

\section{Acceleration Model\label{sec:Acceleration}}

We now return to eq.(\ref{eq:kappa-par}) and rewrite it as follows,
$\kappa_{\parallel}=\kappa_{0}/E_{w}$ with $\kappa_{0}\propto\epsilon^{3/2}$.
To make contact with traditional treatments, at first, we use the
distribution function with the particle density normalized to $fp^{2}dp$
and will return to the particle flux $F\left(\epsilon\right)$ for
a comparison with the data later. The stationary convection-diffusion
equation for $f$ at the shock front has the following form:

\begin{equation}
u\frac{\partial f}{\partial z}=\frac{\partial}{\partial z}\kappa_{\parallel}\cos^{2}\vartheta_{Bn}\frac{\partial f}{\partial z}-\frac{p}{3}\Delta u\frac{\partial f}{\partial p}\delta\left(z\right)+Q\left(p\right)\delta\left(z\right)\label{eq:DC1}
\end{equation}
The last two terms on the r.h.s. are associated with the particle
acceleration on the velocity jump $\Delta u\equiv u-u_{\text{d}}$,
preceded by their injection at the rate $Q$ from a thermal plasma
core \citep{MDru01}. Here $u$ is the upstream flow speed, as before,
and $u_{d}$- the downstream flow speed, both measured in the shock
frame.

Let us integrate eq.(\ref{eq:DC1}) with $\kappa_{\parallel}=\kappa_{0}/E_{w}$
in the upstream region ($z<0$, two last terms on the r.h.s. dropped),
at first once:

\begin{equation}
E_{w}^{-1}\cos^{2}\vartheta_{Bn}\kappa_{0}\frac{\partial f}{\partial z}-uf\left(z\right)=\Phi\left(p\right)\label{eq:FiOfP1}
\end{equation}
The integration constant $\Phi\left(p\right)$ is thus minus the total
$z$- independent particle flux (diffusive plus convective) on the
l.h.s. If $\Phi$ is known, we can determine the particle distribution
at the shock, $f_{0}\left(p\right)$, which is observed to decay approximately
as $p^{-4}$, as in strong shocks. The particle distribution $f\left(p,z\right)$
flattens upstream to $p^{-2}$. Based on the observations shown in
Fig.\ref{fig:ObservSpectra}, the flattening occurs over a short distance
where the particle intensity declines progressively more steeply in
the upstream direction as the particle energy decreases. In the traditional
DSA framework, $\Phi\left(p\right)$ can be specified using the far-upstream
convective flux $\Phi=\Phi_{\infty}=$$-uf\left(-\infty,p\right)\equiv-uf_{\infty}\left(p\right)$,
provided that $\partial f/\partial z\to0$, $E_{\text{w}}\neq0$ at
$z\to-\infty$. However, given the data shown in Fig.\ref{fig:ObservSpectra},
there are problems with this identification of $\Phi\left(p\right)$,
which we discuss below.

A decaying particle flux upstream in Fig.\ref{fig:ObservSpectra}
abruptly changes to a constant $f_{\infty}\left(p\right)$ in all
energy channels at $z=-z_{\infty}\left(p\right).$ The value of $z_{\infty}$
is larger, and the transition is sharper at higher particle energies.
With a good accuracy, the derivative $\partial f/\partial z$ can
be regarded discontinuous at $z=-z_{\infty}$. Hence, in the context
of convection-diffusion problem formulated in $z\in\left(-\infty,\infty\right)$,
the break at $z=-z_{\infty}$ in the particle spectrum (jump of $\partial f/\partial z$)
requires an extra term $-S\left(p\right)\delta\left(z+z_{\infty}\right)$
on the r.h.s. of eq.(\ref{eq:DC1}). It effectively represents a particle
sink, similar to the source of injected particles $+Q\delta\left(z\right)$,
but with an opposite sign. Physically, it means that upon diffusing
to the point $z=-z_{\infty}$ against the plasma flow, particles injected
and accelerated at the shock front earlier promptly escape toward
$-\infty$.

It is worthwhile to compare the boundary $z_{\infty}$ with a so-called
free-escape boundary (FEB), broadly used in Monte Carlo simulations
of the DSA, e.g., \citep{Ellison1990}. There is a significant physical
difference between the $z_{\infty}$ and FEB. Namely, $z_{\infty}$
depends strongly on $p$, while the FEB in simulations typically does
not. Another practical observation is that $z_{\infty}\left(p\right)$
is primarily defined by the background spectrum $f_{\infty}\left(p\right)$,
at least for the flat particle fluxes shown.

Since the particle distribution is constant outside of the interval
$\left(-z_{\infty},0\right)$, it is plausible to formulate the boundary
value problem for eq.(\ref{eq:DC1}) in the finite interval $z\in(-z_{\infty},0)$
upstream instead of $z\in(-\infty,0)$, which would be typical for
the traditional DSA. We, therefore, must set $f=f_{\infty}\left(p\right)$
as the left boundary condition at $z=-z_{\infty}$. Both $f_{\infty}$
and $z_{\infty}$ can be extracted from the data in all energy channels
to calibrate the acceleration model in the next section.

As a second boundary condition for eq.(\ref{eq:DC1}), it is natural
to set $f=f_{0}\left(p\right)$ at $z=0$. We will argue that it is
also worth extracting from the data rather than calculating it using
shock parameters. The difference with the left boundary condition
is that $f_{0}$ can, ideally, be obtained ``from the first principles''
by integrating eq.(\ref{eq:DC1}) across $z=0$, which yields the
following differential equation for $f_{0}$:

\begin{equation}
\Phi\left(p\right)=-uf_{0}-\frac{p}{3}\Delta u\frac{\partial f_{0}}{\partial p}+Q\left(p\right)\equiv\Phi_{0},\label{eq:FiOFp2}
\end{equation}
where $\Phi\left(p\right)$ is the same integration constant in eq.(\ref{eq:FiOfP1}).
It is instructive to return for a moment to the traditional DSA in
which one sets $\Phi_{0}=\Phi_{\infty}=-uf_{\infty}\left(p\right)$.
By substituting this relation into eq.(\ref{eq:FiOFp2}), one then obtains
the solution $f_{0}\left(p\right)$ at and behind the shock. However,
this approach to obtaining the integration constant $\Phi\left(p\right)$
in eq.(\ref{eq:FiOfP1}) has the following problems.

First, the accuracy of eqs.(\ref{eq:DC1}) and (\ref{eq:FiOfP1})
is questionable near the shock since the quasilinear expression for
the diffusive flux is not a good approximation where the wave intensity
is high, and the particle intensity changes sharply. Second, the shock
speed parameters $u$ and $u_{\text{d}}\equiv u-\Delta u$ entering
eq.(\ref{eq:FiOFp2}) significantly fluctuate near the shock transition.
For example, the authors of \citep{Perri2023} indicate a strong deviation
for the shock compression, $r\equiv u/u_{\text{d}}=3.0\pm0.6$. So,
this parameter does not define the spectrum accurately. Thirdly,
we have neglected the cross-field transport, which will likely be
significant near the discontinuity. Lastly, available analytic calculations
of the injection rate $Q\left(p\right)$, which is needed for computing
$f_{0}\left(p\right)$ in eq.(\ref{eq:FiOFp2}), still require a seed
particle source at $p\sim mu$ \citep{MDru01}. The simulations are
generally helpful, but they still have significant disagreements.
They become particularly evident when comparing particle injection
into the DSA with different mass-to-charge ratios (A/Q) (see
\cite{Hanusch2019ApJ} and references therein). This paper scrutinizes numerical and analytical injection models using their predictions of A/Q injection patterns of different ions. Indeed, trajectories of nonrelativistic particles in electromagnetic fields are differentiated by this ratio, as opposed to the ultrarelativistic particles for which the rigidity defines trajectory.    

In light of these problems, we extract $f_{0}\left(p\right)$
for a boundary condition in eq.(\ref{eq:FiOfP1}) at $z=0$, \emph{directly}
from the data. It provides a more stable time-averaged boundary value
than we could possibly calculate using eq.\ref{eq:FiOFp2}), given
the uncertainties listed above. Within this formulation, we will at
first, regard the total particle flux across the shock precursor,
$\Phi$, as a free parameter that will be determined from the boundary
condition at $z=-z_{\infty}$, at which it transitions into the particle
flux escaping the shock to $z=-\infty$.

\section{Solving and Calibrating the Acceleration Model \label{sec:Fitting-the-Spectra}}

After expressing the wave energy density $E_{w}\left(z,\epsilon\right)$
in eq.(\ref{eq:FiOfP1}) and the particle distribution $f$ through
the particle flux $F$, using eq.(\ref{eq:Ew-of-P}) and the relation
$F\left(\epsilon\right)d\epsilon=f\left(p\right)vp^{2}dp$, respectively,
we evaluate eq.(\ref{eq:FiOfP1}) to the following form:
\begin{equation}
K\left(z\right)\frac{\kappa_{0}\left(\epsilon\right)\epsilon^{-3/2}}{F+\psi\left(\epsilon\right)}\frac{\partial F}{\partial z}-F=\Psi\left(\epsilon\right),\label{eq:F-final}
\end{equation}
where we have introduced the notation
\begin{equation}
K\equiv\frac{3\rho V_{A}\left(1-M_{A}^{-1}\right)}{16\pi\sqrt{2m}}\cos^{2}\vartheta_{\text{Bn}}\left(z\right);\;\;\Psi\left(\epsilon\right)=\frac{2m}{u}\epsilon\Phi.\label{eq:K-PsiDef}
\end{equation}
Recall that $\psi$ and $\Psi$ emerged from the integration of the
wave production balance in eq.(\ref{eq:WaveGen}) and that of the
convective-diffusive particle transport in eq.(\ref{eq:DC1}), respectively.
As we discussed in the previous section, $\Psi$ can formally be expressed
through the particle injection rate $Q$ and the upstream and downstream
flow speeds $u$ and $u_{d}$, using eq.(\ref{eq:FiOFp2}). By omitting
$Q$ for nonthermal particles, for $\Psi$ we have:
\begin{equation}
\Psi\left(\epsilon\right)=-\frac{\epsilon^{-s+1}}{s+1}\frac{\partial}{\partial\epsilon}\epsilon^{s}F_{0}.\label{eq:PsiVsF}
\end{equation}
Here, we have converted the power-law index $q$, introduced earlier
for $f_{0}\left(p\right)$, to its equivalent for $F_{0}=F\left(\epsilon,z=0\right)$:
\begin{equation}
s=\frac{3u}{2\Delta u}-1=\frac{q}{2}-1.\label{eq:sIndDef}
\end{equation}
Although the contribution of injection term $Q$ is negligible at
high energies, it defines the normalization of $F_{0}$, and thus
$\Psi$. However, $\Psi\left(\epsilon\right)$ vanishes where $F_{0}$
follows the DSA spectrum $\propto\epsilon^{-s}$, but as $s$ is known
only approximately, we keep $\Psi$ in the analysis. Since we use
a steady-state model, while the shock parameters likely fluctuate,
it is difficult to calculate $\Psi$ \emph{a priori}. A more plausible
approach is to extract it by fitting the solution of eq.(\ref{eq:F-final})
to the data, although we will constrain $\Psi$ below.

The measured downstream spectrum closely follows the DSA predictions
if one takes the uncertain compression ratio close or slightly above
the upper bound given by \citep{Perri2023}, $r=3.6$. It means that
$\Psi$ is small compared to $F_{0}$ and even $F\left(z\right)$
in a significant part of the shock precursor, $\Psi\ll F$. Other
than that, $\Psi$ is largely unconstrained, as long as we do not specify the 
injection efficiency and shock compression. To minimize the number
of parameters it depends upon, we will represent $\Psi$ by a power-law
function of $\epsilon$ with an amplitude and index extracted from
the data rather than precalculated from the relation in eq.(\ref{eq:PsiVsF}).

Turning to the parameter $\psi$, from Fig.\ref{fig:ObservSpectra}
we see that near $z=-z_{\infty}$, $F$ is way below its values almost
across the entire shock precursor. By eq.(\ref{eq:WaveGen}), the
same conclusion can be drawn for $E_{w}$. Hence, $\psi\left(p\right)$
in eq.(\ref{eq:Ew-of-P}) must also be small compared to $F$, except
near $z=-z_{\infty}$. Summarizing the above considerations, we treat
$\Psi$ and $\psi$ as free parameters and use $F_{0}$ and $F_{\infty}$
as more reliable inputs for the model since we determine them directly
from the data. However, we have also used them to constrain the less
certain parameters $\Psi$ and $\psi$.

\begin{figure}
	\centering
	\includegraphics[width=1.0\columnwidth]{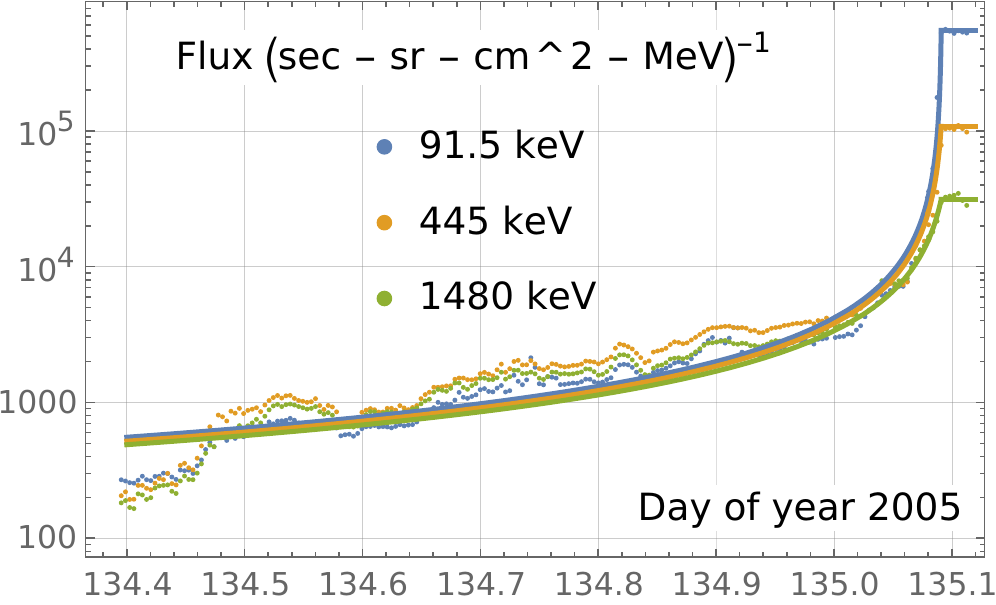}
	\caption{Approximate solution in eq.(\ref{eq:FsolAppr2}) shown in three energy
		channels indicated in the plot. \label{fig:Approximate-solution-eq.()}}
\end{figure}%

After introducing a normalized distance $\zeta$ in place of $z$,
which has the dimensionality of $1/F$,
\begin{equation}
\zeta=\frac{\epsilon^{3/2}}{\kappa_{0}\left(\epsilon\right)}\int_{0}^{z}\frac{dz^{\prime}}{K\left(z^{\prime}\right)},\label{eq:zeta-Def}
\end{equation}
and rewriting eq.(\ref{eq:F-final}) as
\begin{equation}
\frac{1}{F+\psi}\frac{\partial F}{\partial\zeta}-F=\Psi,\label{eq:F-final-ND}
\end{equation}
we obtain its solution, that satisfies the boundary condition $F\left(0,\epsilon\right)=F_{0}\left(\epsilon\right)$,
for arbitrary values of parameters $\psi$ and $\Psi$:
\begin{equation}
F=\left(\Psi-\psi\right)\left[\left(1+\frac{\Psi-\psi}{F_{0}+\psi}\right)e^{\left(\psi-\Psi\right)\zeta}-1\right]^{-1}-\psi.\label{eq:Fexact}
\end{equation}
The solution at large negative $\zeta$, which we associate with an
area of far upstream transition to the background population of energetic
particles, depends on $\Psi$ and $\psi$. Remarkably, in the near
upstream area of transition from the downstream spectrum $F_{0}\left(\epsilon\right)$
to the broad  flat-flux area, including the latter, 
the solution is insensitive to these parameters. It is, therefore,
worthwhile to begin the solution analysis with small negative $\zeta$
since it helps understand how this transition from the downstream
$F_{0}\left(\epsilon\right)$ to the flat upstream distribution occurs.

\subsection{Near- and Mid-Upstream Zones\label{subsec:Near-upstream}}
\begin{table*}
	\centering
	\hspace{-12em}
	\begin{tabular}{|c|c|c|c|c|c|c|c|c|c|c|c|c|c|c|}
		\hline 
		& \multicolumn{4}{c|}{Function of $\epsilon$(keV)} & \multicolumn{10}{c|}{Parameters}\tabularnewline
		\hline 
		-  & $F_{0}$  & $\Psi$  & $\psi$  & $\zeta$  & $C_{0}$  & $C_{\Psi}$  & $C_{\psi}$  & $C_{\zeta}$  & $\sigma$  & $\gamma$  & $\rho$  & $\mu$  & $\epsilon_{1}$  & \multicolumn{1}{c|}{$t_{\text{sh}}$}\tabularnewline
		\hline 
		\begin{turn}{0}{Fit}
		\end{turn}  & \begin{turn}{90}
			$C_{0}\left(\epsilon_{1}/\epsilon\right)^{\sigma}$ 
		\end{turn} & \begin{turn}{90}
			$C_{\Psi}\left(\epsilon_{1}/\epsilon\right)^{\gamma}$ 
		\end{turn} & \begin{turn}{90}
			$C_{\psi}\left(\epsilon_{1}/\epsilon\right)^{\rho}$ 
		\end{turn} & \begin{turn}{90}
			$C_{\zeta}\left(\epsilon/\epsilon_{1}\right)^{\mu}\left(t-t_{\text{sh}}\right)$ 
		\end{turn} & \begin{turn}{90}
			5.5$\times10^{5}$ 
		\end{turn} & \begin{turn}{90}
			685 
		\end{turn} & \begin{turn}{90}
			185 
		\end{turn} & \begin{turn}{90}
			$2.61\times10^{-3}$ 
		\end{turn} & \begin{turn}{90}
			1.03 
		\end{turn} & \begin{turn}{90}
			0.103 
		\end{turn} & \begin{turn}{90}
			0.14 
		\end{turn} & \begin{turn}{90}
			0.04 
		\end{turn} & \begin{turn}{90}
			91.5 keV 
		\end{turn} & \begin{turn}{90}
			135.091 (DoY) 
		\end{turn}\tabularnewline
		\hline 
		\begin{turn}{0}{Eq./BC} \end{turn} & BC  & \ref{eq:PsiVsF}  & \ref{eq:Ew-of-P}  & \ref{eq:zeta-Def}  &  &  &  &  &  &  &  &  &  & \tabularnewline
		\hline 
	\end{tabular}
	\caption{Input functions and parameters for eq.(\ref{eq:Fexact}). The downstream
		normalization constant, $C_{0}$, and the power-law index $\sigma$
		are extracted directly from the spectrum downstream. Other parameters
		are obtained by fitting the data, including the breaking points at
		$z=-z_{\infty}$ to eq.(\ref{eq:Fexact}), shown in Fig.\ref{fig:Fits}.
		(see text). \label{tab:Input-functions}}
\end{table*}

By developing the general solution $F$ in eq.(\ref{eq:Fexact}) in
a Taylor series in $\left|\left(\psi-\Psi\right)\zeta\right|\ll1$
and assuming, as argued earlier, that $\psi\ll F$ in the most of
the shock precursor, we find that within these approximations $F$
does not depend on $\Psi$ or $\psi$:

\begin{equation}
F\simeq\frac{F_{0}}{1-F_{0}\zeta}.\label{eq:FsolAppr2}
\end{equation}
Using the data, we fit $F_{0}\left(\varepsilon\right)$ by $F_{0}\propto\epsilon^{-\sigma}$,
where $\sigma\approx1.03$, Table \ref{tab:Input-functions}. To convert
the normalized distance to the shock, $\zeta$, to the spacecraft
time, we first relate the latter to the physical distance as follows:
$z=U_{\text{sc}}\left(t-t_{\text{sh}}\right)$. 
Here, $U_{\text{sc}}$
denotes the spacecraft speed relative to its shock crossing point
at the time instance $t_{\text{sh}}$. This conversion thus combines this geometrical  uncertainty with the $\kappa_{0}$ parametrization in eq.(\ref{eq:zeta-Def}), we include using eq.(\ref{eq:kappa-par}) as follows
\[
\kappa_{0}\approx\left(\frac{\epsilon}{\epsilon_{1}}\right)^{3/2}\frac{v_{1}^{2}}{\omega_{\text {ci}}}w,  \; {\text{where}}\; w=\frac{v_{1}}{l\omega_{\text {ci}}}\left(\frac{B_{0}}{\delta B_{l}}\right)^{2},
\]
where $\epsilon_{1}=91.5$ keV is the particle energy in the lowest
energy channel that we use in our analysis, $v_{1}\approx3\times10^{8}$cm/s
is the respective particle velocity, and $l$ is the dominant scale
of nonresonant magnetic perturbations responsible for particle scattering.
We will discuss other aspects of the mapping $z\to\zeta$ when presenting
the full fit of the solution in eq.(\ref{eq:Fexact}) to the data
shown in Fig.\ref{fig:ObservSpectra} in the next subsection.

We plot the formula in eq.(\ref{eq:FsolAppr2}) against time and compare
the result with the data in Fig.\ref{fig:Approximate-solution-eq.()}.
For clarity, we have selected three representative energies from the
six shown in Fig.\ref{fig:ObservSpectra}. Apart from a good agreement
with the data, the simplified expression in eq.(\ref{eq:FsolAppr2})
does not include uncertain model parameters $\Psi$ and $\psi$. This
expression thus embodies the universality of the underlying flattening
mechanism. The model parameters are lumped here in a single
variable, $\zeta$, defined in eq.(\ref{eq:zeta-Def}). They include
a combination of the wave turbulence level, $w$, its characteristic
scale, $l$, most vital for the particle scattering, the local field
angle $\vartheta_{nB}\left(z\right)$, plasma density,  magnetic
field strength, and the spacecraft trajectory angle relative to the shock surface, mentioned above. Some of these quantities are uncertain and even fluctuate, 
but we combine them in  a single constant
$C_{\zeta}$ in Table \ref{tab:Input-functions} to describe this
multi-variable combination. For a better agreement with the data far
upstream, considered below, we have also included a small
correction to the $\kappa_{0}\propto\epsilon^{3/2}$ scaling by representing it as $\epsilon^{3/2-\mu}$,
with $\mu=0.04$.

When comparing the model prediction with the data shown in Fig.\ref{fig:ObservSpectra},
we note that they provide only a single-pass scan of the particle
intensity with no direct information about their possible time variability.
The DSA acceleration time, however, is typically the shock-crossing
time of its precursor filled with the accelerated particles, $\tau_{\text{acc}}\sim\kappa/U_{\text{sh}}^{2}$.
Although this estimate is strictly applicable to a linear acceleration
regime with a prescribed, flux-independent $\kappa\left(\epsilon\right)$,
it can be shown to apply also to strongly nonlinear particle acceleration
in shocks modified by their pressure \citep{MDru01}. The case we
consider belongs to neither of the above. However, the acceleration
time is much longer than the shock parameter variation time scale,
which can be inferred from the short-scale variation of the plasma
and particle data across the precursor. As our model is stationary,
the deviations of the real data from the theoretical curve are expected.
The agreement may be partly improved by including the $z$ dependence
of $\vartheta_{nB}$ in the definition of $\zeta$. However, the nonmonotonic
parts of the particle intensity cannot be corrected within the steady-state
model. We will return to the observed deviation in the Discussion
section.

Since $\Psi,\psi\ll F_{0}$, we can further simplify the expression
in eq.(\ref{eq:FsolAppr2}) for larger $-\zeta$ without violating
its applicability condition, $\left|\left(\psi-\Psi\right)\zeta\right|\ll1$.
In this approximation, the upstream spectrum $F\left(\zeta\right)$,
$\zeta<0$ ceases to depend on $\epsilon$ \emph{explicitly.} The
transition to this regime occurs over a narrow interval adjacent to
the shock, $-1/F_{0}\lesssim\zeta<0$, where the range of $1/F_{0}$
is roughly between $2\times10^{-6}<1/F_{0}\left(\epsilon\right)<3\times10^{-5}$.
To the left of this interval, i.e., for $\zeta<-1/F_{0}$, the spectrum
depends only on $\zeta$ and can be approximated as

\begin{equation}
F\approx-1/\zeta,\label{eq:OneOverZeta}
\end{equation}
up to the far upstream, where it gradually decreases to $F\sim\psi$
or $F\sim\Psi$, whichever occurs closer to the shock. In this universal
$1/\zeta$ -regime, the energy dependence may only be through the
normalized distance $\zeta$, defined in eq.(\ref{eq:zeta-Def})
and discussed at some length above. As we have shown, for a nonresonant
particle scattering the particle diffusivity must scale close to $\kappa_{0}\left(\epsilon\right)\propto\epsilon^{3/2}$,
eq.(\ref{eq:kappa-par}). Therefore, $\zeta$ does not depend on $\epsilon$
and the asymptotic solution $F\approx-1/\zeta$ is, indeed, perfectly
flat.

Most of the spectrum is thus describable by a simple formula in eq.(\ref{eq:FsolAppr2}),
simplifying even further beyond a narrow layer upstream of the shock
surface. A salient feature of this solution is that the uncertain model parameters
$\psi\left(\epsilon\right)$ and $\Psi\left(\epsilon\right)$ do not
enter it. All essential aspects of this solution are encapsulated
in the single variable $\zeta$. Yet, it agrees with the data, as
shown in Fig.\ref{fig:Approximate-solution-eq.()}. However, some
deviations are present. In the Discussion section, we will consider
three probable causes: time variability of the shock parameters and
particle acceleration, magnetic particle traps upstream, and the coordinate
dependence of the shock angle.

\subsection{Far Upstream Zone\label{subsec:Far-Upstream-Zone}}

To extend the fit shown in Fig.\ref{fig:Approximate-solution-eq.()}
to the outermost part of the upstream plasma, we need to specify the
parameters $\psi$ and $\Psi$ of the full solution in eq.(\ref{eq:Fexact}).
We argue below that these parameters are indeterminate within the
given data set and should be treated as free model parameters. Nevertheless,
they can be constrained within our model beyond the conditions $\Psi,\psi\ll F_{0}$, mentioned earlier.

First of all, they emerged as integration constants for a \emph{steady
state} solution of the acceleration problem. If the steady state were
absolute, we would be able to calculate the parameter $\Psi$ using
eq.(\ref{eq:PsiVsF}): $\Psi=-\left(s-\sigma\right)F_{0}\left(\epsilon\right)/\left(s+1\right).$
On the other hand, taking the (steady state) DSA prediction at its
face value, we should expect $s=\sigma$, meaning $\Psi\to0$ for
sufficiently large $\epsilon$ where the injection from the thermal
plasma core given by $Q\left(p\right)$ in eq.(\ref{eq:DC1}) fades
out, irrespective of the seed particles for injection \citep{MDru01}.
Clearly, the $\Psi=0$ condition constitutes an exact balance between
the convective and diffusive particle fluxes in the shock precursor,
cf. eq.(\ref{eq:F-final-ND}), resulting from the steady state assumption.
As we argued earlier, it is violated in a realistic time-dependent
situation.

An indication of time variability of the data provides an independently
measured shock compression, $r$, yielding a steady state DSA index
$s=(r+2)/2(r-1)$ that does not match the index $\sigma$, directly
obtained from the downstream spectrum. The shock compression ratio
estimated in \citep{Perri2023} in the range $2.4<r<3.6$ maps to
the range of $s=1.08-1.57$, whereas the measured downstream spectrum
is outside of that range, $\sigma=$1.03. It should be noted, though,
that the shock may be modified by accelerated particles, which is
visible in some observations, e.g., \citep{TerasawaHelio11}. Such
a modification would increase the total shock compression between
the far upstream and downstream flow but decrease the shock compression
at the flow discontinuity \citep{MDru01}, often termed ``subshock''.
Nevertheless, the flow modification would also violate the above relation
between $\Psi$ and $F_{0}$ because of a particle acceleration term,
$\propto du/dz$, in the shock precursor.

\begin{figure*}
	\centering
	\includegraphics[scale=0.58]{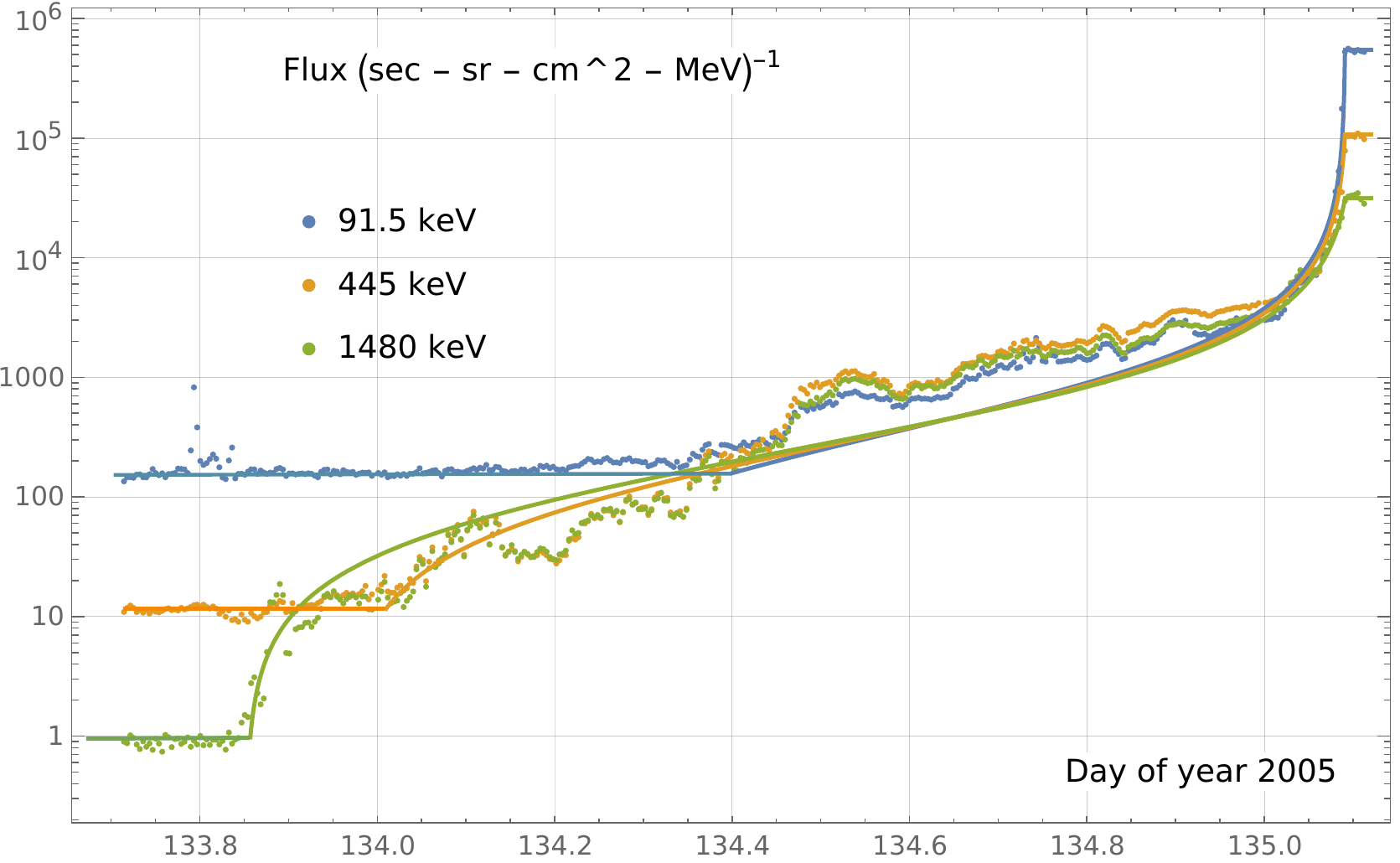}
	\caption{Fits in three energy channels produced using eq.(\ref{eq:Fexact})
		with parameters listed in Table \ref{tab:Input-functions}.\label{fig:Fits}}
\end{figure*}

Finally, the above relation between $\Psi$ and $F_{0}$ is obtained
by considering the flux balance at the shock transition. As we have
seen, in the upstream region adjacent to the shock, the solution is
not sensitive to $\Psi$, and a directly measured $F_{0}$ is sufficient
to consistently describe the spectrum in the near and mid-upstream
areas. It seems then reasonable not to bind the constant $\Psi$ to
$F_{0}$ by the above relation, inferred from the solution near the
shock, especially because $\Psi$ significantly affects the solution
on the opposite end of the flat spectrum, which is far upstream.

Similar arguments apply to $\psi$, introduced in eq.(\ref{eq:Ew-of-P}).
The flux $F$ varies by several orders of magnitude between the shock
and far upstream region, and only in the latter $\psi$ affects the
result, while being negligible otherwise. Even the sign of $\psi$
can be positive or negative depending on the relation between the
background particle and wave energy density far upstream, just like
in the case of $\Psi$. Some simple physical arguments are in order
upon their signs.

Measured in the shock frame, the positive total particle flux $\Psi$
in eq.(\ref{eq:F-final-ND}) means that the shock-accelerated particles
are transported away from it in the upstream direction, and their
flux is predominantly diffusive. It also means that the accelerated
particles are copiously injected at the shock instead of being accelerated out
of the preexisting background population far upstream. The $\psi>0$
condition, in turn, indicates that the background turbulence energy
density far upstream dominates that of the energetic particles. The
reverse arguments also apply in both instances. It seems that the
only firm constraint that we can impose on these quantities is, indeed,
$\Psi,\psi\ll F_{0}$.

That said, the sign of $\chi\equiv\Psi-\psi$ is crucial to the asymptotic
spectra far upstream. From eq.(\ref{eq:Fexact}) we deduce that $F\to-\psi$
for $\chi>0$ and $F\to-\Psi$ for $\chi<0$, when $\zeta\left|\chi\right|\to-\infty$.
We can thus formally satisfy the boundary condition $F\to F_{\infty}\left(\epsilon\right)$
by choosing either the integration constant $\psi=-F_{\infty}$ or
$\Psi=-F_{\infty},$ depending on whether $\chi=\Psi-\psi$ is positive
or negative. These choices satisfy the boundary condition at $z=-\infty$,
but approaching $F_{\infty}$ is then gradual. We see from Fig.\ref{fig:ObservSpectra},
however, that $F\left(z\right)$ abruptly turns to an energy-dependent
constant $F_{\infty}\left(\epsilon\right)$ at $z=-z_{\infty}\left(\epsilon\right)$.
The spectrum ceases to be flat beyond this point. This abrupt transition
requires the boundary condition $F\left(z\le-z_{\infty}\right)=F_{\infty}$,
with a break in $z$- derivative at $z=-z_{\infty}$, as mentioned
earlier. So, we have two input parameters $F_{\infty},z_{\infty}$
instead of one $F_{\infty}$. By extracting them for each particle
energy from the data, we determine the two unknown parameters, $\Psi$,
and $\psi$, entering the full solution in eq.(\ref{eq:Fexact}).

To summarize our spectrum fitting procedures,
the normalized coordinate $\zeta$  contains the combination $\epsilon^{3/2}/\kappa_{0}\left(\epsilon\right)$
that we argued to be $\epsilon$- independent. This is an accurate
but not exact statement. We, therefore, introduce a weak
power-law dependence of this quantity, $\propto\epsilon^{0.04}$,
for a better agreement of the breaking points, $z_{\infty}\left(\epsilon\right)$,
in all energy channels. Likewise, we specify the functions $\Psi\left(\epsilon\right)$
and $\psi\left(\epsilon\right)$ as power law functions. These quantities
are defined by constant power law indices ($\mu,\gamma$, and $\rho$
in Table \ref{tab:Input-functions}). As we fit a continuum of profiles
$F\left(\epsilon,z\right)$ using these three indices, the fine-tuning
concerns do not apply. The fits are shown in Fig.\ref{fig:Fits},
and the required parameters are summarized in Table \ref{tab:Input-functions}.

\section{Conclusions}

The following two modifications to the diffusive shock acceleration
(DSA) theory suffice to explain spectral flattening observed ahead
of several interplanetary shocks: 
\begin{itemize}
\item Inclusion of a dependence of particle diffusivity $\kappa$ on the
particle flux $F$ (nonlinear particle transport) that, in turn, is
directly related to the scattering wave intensity 
\item Switching from the traditional DSA resonant wave-particle interaction
for short-scale magnetic perturbations that are also self-consistently
generated by, but not resonant with, accelerated particles 
\end{itemize}
In the resulting DSA solution, the nonresonant, nonlinear particle
diffusivity, $\kappa$, increases with energy as $\propto\epsilon^{3/2}$,
simultaneously decreasing with the wave energy as $E_{w}^{-1}\propto\epsilon^{-3/2}F^{-1}$,
thus turning not explicitly depending on the particle energy almost
everywhere in the shock precursor. This independence results in a
diffusive flux $\propto$$F^{-1}\partial F/\partial z$ that being
not energy dependent explicitly and balanced with the convective flux,
$uF$, results in an energy-independent $F\left(z\right)$.

\section{Discussion\label{sec:DiscOfRes}}

The presented acceleration model reproduces a surprisingly flat particle
spectrum observed upstream of an interplanetary shock, including its
transitions to a regular diffusively accelerated particle spectrum
downstream and the inflowing background spectrum far upstream. Meanwhile,
some spatial deviations of the flat part of the spectrum from the
data remain. In part, they can be explained by variations in the shock angle entering the 
normalized coordinate $\zeta$ in eq.(\ref{eq:zeta-Def}).
However, their nonmonotonic parts cannot be removed within our steady-state
model since the deviations are almost certainly time-dependent. This
can be seen from the sign changes in the spatial gradient of the flux
data upstream. If $\partial F/\partial z<0$, both diffusive and convective
fluxes are directed to the shock and cannot cancel out, thus precluding
a steady-state solution in the shock frame. At the same time, nonmonotonic
deviations from the predicted profile may constitute bunches of particles
trapped in traveling magnetic disturbances. For example, they may
result from magnetic perturbations driven by accelerated particles,
subsequently steepening into shocklets or shocktrains upstream (see,
e.g., \cite{Kennel_MSSK_88,MD09} and references therein).

More broadly, magnetically trapped particles may originate from intrinsic
shock instabilities, such as shock reformation and shock corrugations \citep{Burgess12,CaprSpitk14a},
that result in an impulsive release of accelerated particles upstream.
They propagate then in bunches away from the shock. They may also
be trapped in magnetic bubbles, self-created, or preexisting in the
solar wind. Since the accelerated particles near the shock upstream
typically have an anisotropic distribution, they may drive mirror
and firehose instabilities, resulting in magnetic bubbles that trap
energetic particles. Consistent descriptions of their dynamics require
a significant model extension, including a time-dependent shock description
beyond one dimension.

The role of magnetic bubbles upstream and their interaction with energetic
particles have already been discussed in an analysis of a 1978 interplanetary
shock in \citep{Kennel1986}. These authors presented a detailed data
comparison with the Lee theory \citep{Lee1983}, reaching a much closer
agreement between the two than we found in the 2005 shock considered
in the present paper. In particular, no spectrum flattening was observed
in the 1978 shock, which is in complete agreement with Lee's predictions.
The difference between the two shocks is that the 1978 shock has a
significantly steeper downstream spectrum in the range $q=4.20-4.25$,
while the 2005 shock has $q\approx4.06$ pointing to a considerably
higher shock compression. However, the compression ratio estimate
of \citep{Perri2023}, $r\approx3$, formally results in $q\approx4.5$.
The reason for this disagreement is, at least in part, that the above
estimate of the shock compression has likely been obtained by analyzing
the flow density and speed immediately upstream and downstream of
the discontinuity. By inspecting these flow characteristics further
upstream in Fig.1 of the above paper, one may see that the shock is
significantly modified, most likely by accelerated particles penetrating
upstream. The total compression is close to four, consistent with
the particle spectral index used in our paper.

\vspace{1em}
\noindent
\nolinenumbers
This research work is supported by the NASA grant 80HQTR21T0005. MM
gratefully acknowledges additional support from the NSF grant AST-2109103.

\bibliographystyle{aasjournal}
\bibliography{}

\end{document}